\documentclass[12pt]{iopart}

\usepackage{graphicx}
\usepackage{psfrag}
\def\la{~\mbox{\raisebox{-.6ex}{$\stackrel{<}{\sim}$}}~}
\def\ga{~\mbox{\raisebox{-.6ex}{$\stackrel{>}{\sim}$}}~}
\def\beq{\begin{equation}}
\def\eeq{\end{equation}}

\begin{document}

\title[No-Scale Inflation]{No-Scale Inflation}

\author{John Ellis}

\address{Department of Physics, King's College London, London WC2R 2LS, UK; \\
Theory Division, CERN, CH-1211 Geneva 23, Switzerland}
\ead{John.Ellis@cern.ch}

\author{Marcos A. G. Garcia}

\address{William I. Fine Theoretical Physics Institute, School of Physics and Astronomy, University of Minnesota, Minneapolis, MN 55455, USA}
\ead{garciagarcia@physics.umn.edu}

\author{Dimitri V. Nanopoulos}

\address{George P. and Cynthia W. Mitchell Institute for Fundamental Physics and Astronomy, Texas A\&M University, College Station, TX 77843, USA; Astroparticle Physics Group, Houston Advanced Research Center (HARC), Mitchell Campus, Woodlands, TX 77381, USA;
Academy of Athens, Division of Natural Sciences, Athens 10679, Greece}
\ead{dimitri@physics.tamu.edu}

\author{Keith A. Olive}

\address{William I. Fine Theoretical Physics Institute, School of Physics and Astronomy, University of Minnesota, Minneapolis, MN 55455, USA}
\ead{olive@physics.umn.edu}

\vspace{10pt}
\begin{indented}
\item[]July 2015
\end{indented}

\begin{abstract}
Supersymmetry is the most natural framework for physics above the TeV scale, and
the corresponding framework for early-Universe cosmology, including inflation, is supergravity. No-scale supergravity
emerges from generic string compactifications and yields a non-negative potential, and is therefore a plausible framework
for constructing models of inflation. No-scale inflation yields naturally predictions similar
to those of the Starobinsky model based on $R + R^2$ gravity, with a tilted
spectrum of scalar perturbations: $n_s \sim 0.96$, and small values of the tensor-to-scalar
perturbation ratio $r < 0.1$, as favoured by Planck and other data on the cosmic microwave background (CMB).
Detailed measurements of the CMB may provide insights into the embedding of inflation within string theory
as well as its links to collider physics.
\end{abstract}
\begin{flushright}
{\small KCL-PH-TH/2015-28, LCTS/2015-20, CERN-PH-TH/2015-144 \\ 
ACT-05-15, MI-TH-1521, UMN-TH-3442/15, FTPI-MINN-15/32}
\end{flushright}
%
%
%
%

%

\section{Introduction}

Data on the cosmic microwave background (CMB) from the Planck satellite~\cite{Planck15}
and other experiments are qualitatively consistent with generic expectations from cosmological
inflation~\cite{Inflation}. In particular, they are consistent with the prototypical Starobinsky model based
an $R + R^2$ extension of minimal Einstein gravity~\cite{Staro}, and are constraining or excluding many 
other models. As discussed elsewhere in this volume, 
some still consider alternatives to inflation~\cite{ST}, but we are encouraged by the progressive and impressive
improvement in experimental precision to explore specific Planck-compatible models in more detail.
As we discuss, detailed measurements of the CMB provide a window on fundamental physics that is complementary
to laboratory experiments, casting light on particle physics at energies far beyond the reach of colliders
and possibly giving us insight into string compactifications.

Run~1 of the LHC revealed the Higgs boson~\cite{Higgs}, which is an existence proof for an apparently
elementary scalar boson. As such, it may serve as a prototype for the inflaton, and it has been
proposed that the Higgs could even be the
inflaton itself~\cite{BS}. This requires a rather large non-minimal gravitational coupling of the Higgs field,
and seems impossible unless the Standard Model of particle physics is
supplemented, since naive extrapolation of the Standard Model leads to a negative Higgs potential at high scales.
At the time of writing the LHC has yet to reveal any new physics beyond the Standard Model,
but there are many reasons to expect new physics, and we consider supersymmetry to be the 
best-motivated possibility~\cite{susy}. 

The appearance of the Higgs boson with a mass $\sim 125$~GeV~\cite{125} sharpens the
problem of the naturalness of the electroweak scale, which low-energy supersymmetry could mitigate.
Moreover, simple supersymmetric models actually predicted correctly the mass of the Higgs boson~\cite{ERZ}, and also
that its couplings would resemble those in the Standard Model~\cite{EHOW} - which they do, so far. These are new
motivations for supersymmetry provided by Run~1 of the LHC, in addition to the r\^oles that supersymmetry could play in
grand unified theories and string theory. It is therefore natural to consider supersymmetric models of
inflation.

Supersymmetric versions of inflation were originally proposed in the context of a growing set of problems \cite{olive} besetting
the new inflationary theory \cite{new} based on the one-loop (Coleman-Weinberg) potential for
breaking SU(5). For example, the vacuum tended to evolve to a
minimum different from that containing the Standard Model \cite{vacuum}, and
quantum fluctuations destabilized the inflationary vacuum~\cite{linqf} unless the Higgs effective
quartic self-coupling was small, $\la 10^{-2}$, whereas its value was fixed by the SU(5) gauge coupling
to be $\ga 1$. But the biggest problem for new inflation was the magnitude of density fluctuations \cite{pert}.
The new inflationary model based on SU(5) predicted that they should be ${\cal O}(1)$,
whereas experimentally they are ${\cal O}(10^{-5})$.

Several of these problems are tied to the magnitude of the effective quartic 
potential coupling, which must be tuned to ${\cal O}(10^{-12})$ to insure acceptable density fluctuations.
Indeed, in any model of
inflation based on an elementary scalar field, its effective potential must have some parameter that
is small in natural units where the reduced Planck mass $M_P \equiv 1/\sqrt{8 \pi G_N} \simeq 2.4 \times 10^{18}$~GeV
is set to unity. In a supersymmetric theory, such parameters are renormalized multiplicatively, so the
quantum corrections to small values are under control.

For this reason, it was suggested that inflation cries out for supersymmetry~\cite{Cries}.
In this framework the magnitude of the self-coupling could be linked to the ratio of supersymmetry breaking
to the GUT scale rather than to the GUT gauge coupling alone~\cite{Cries,prim,fluct,NT}. 
Tension due to fine tuning and the duration of inflation could be further relieved
if the inflationary field value were separated from the GUT scale and pushed to the Planck scale,
the scenario of primordial inflation~\cite{prim,prim2} using a gauge singlet field~\cite{prim}
that was baptized the inflaton \cite{nos}.
Primordial supersymmetric inflation made it easy to render natural the fact that the
magnitude of the observed scalar density perturbations is ${\cal O}(10^{-5})$ \cite{fluct}. 

However, it is clear that any discussion of early-Universe cosmology, including inflation, should also incorporate gravity
in an essential way, and hence be set in the framework of supergravity~\cite{sugrainflation,gl}.
In general, a simple supergravity theory
is characterized by a Hermitian function of the matter scalar fields $\phi^i$, called the K\"ahler potential $K$,
that captures its geometry, a holomorphic function of the scalar fields, called the superpotential $W$,
that describes their interactions, and another holomorphic function $f_{\alpha \beta}$ that characterizes their couplings
to gauge fields $V_\alpha$~\cite{sugra}. 

In minimal ${\cal N} = 1$ supergravity, the K\"ahler metric is flat:
\beq
K = \phi^i \phi_i^* \, ,
\eeq
where the sum is over all scalar components in the theory. 
The simplest inflationary theory in minimal supergravity is defined by the superpotential \cite{hrr}
\beq
W = m^2 (1 - \phi)^2 \, ,
\label{hrrW}
\eeq
where $\phi$ is the inflaton and $m \sim 10^{-5}$ in Planck units. However,
because inflation in this model is effectively driven by a cubic term in the scalar potential,
it leads to a prediction for a scalar perturbation spectrum with tilt, $n_s = 0.933$,
which is now in serious disagreement with the determination by Planck \cite{Planck15}: $n_s = 0.968 \pm 0.006$.

Moreover, a generic supergravity theory coupled to matter is not suitable for cosmology,
because its effective scalar potential is proportial to $e^K$, scalars typically pick up masses proportional
to $H^2 \sim V$, where $H$ is the Hubble parameter~\cite{eta}. Though the theory defined by (\ref{hrrW})
is constructed to avoid this $\eta$ problem, a generic inflationary model is in general plagued by this problem of
large masses. In addition, the spontaneous breaking of local supersymmetry introduces additional challenges for constructing a successful
supergravity inflationary scenario \cite{polprob,enq,myy}, stemming from the introduction of a chiral superfield
whose scalar components have weak-scale masses but Planck-scale vacuum expectation values (vevs) \cite{pol,bfs}~\footnote{Some
of these problems are alleviated if the supergravity-breaking field has a parametrically larger mass \cite{enq,myy,jy}
as may occur in theories with strongly-stabilized moduli \cite{dine,Dudas:2006gr,klor,ego}.}.

The question then arises, which type of supergravity theory to choose for formulating models of inflation? 
An attractive way to avoid the $\eta$ problem
is provided~\cite{GMO} by no-scale supergravity~\cite{CFKN,LN}. In the minimal two-field case~\cite{EKNN}
useful for inflation, its K\"ahler potential
can be written in the logarithmic form
\begin{equation}
K \; \ni \; - 3 \, \ln \left(T + T^* - \frac{|\phi|^2}{3}\right) + \dots \, , 
\label{nsK}
\end{equation}
where $T$ and $\phi$ are complex scalar fields and the $\dots$ represent possible additional matter fields~\footnote{The 
$\eta$ problem may also be avoided in theories with a shift symmetry in 
which the K\"ahler potential is a function $(\phi - \phi^*)^2$ rather than $\phi {\phi^*}$ \cite{kyy}. In this case the danergous term arising from $e^K$ does not depend on the combination $\phi + \phi^*$, which may then be chosen as the inflaton.}.
Moreover, no-scale supergravity emerges as the effective four-dimensional low-energy field theory in generic compactifications
of string theory~\cite{Witten}, with $T$ being identified as the compactification volume modulus.
We therefore consider it to be the best-motivated framework for constructing field-theoretical
models of inflation~\cite{GL,KQ,EENOS,otherns,klor}. 

A simple version of a no-scale inflationary model 
is defined by the superpotential $W = m^2 (\phi -\phi^4/4)$~\cite{EENOS}. However, in this model too, inflation is effectively
driven by a cubic term in the scalar potential, so that it
yields the same prediction $n_s = 0.933$ as in the minimal model \cite{hrr}.
The Planck 2013 data~\cite{Planck13}, with their confirmation
of a tilt in the spectrum of scalar perturbations with $n_s \sim 0.96$ and their strengthening of previous
upper bounds on the tensor-to-scalar ratio $r$ triggered to re-examine no-scale inflation~\cite{ENO6}.
As already mentioned, the Planck data are highly consistent with the predictions of
the Starobinsky $R + R^2$ model~\cite{Staro}. We were therefore very impressed to discover that the simplest
possible Wess-Zumino superpotential~\cite{WZ}
\begin{equation}
W(\phi) = \frac{{\hat \mu}^2}{2} \phi^2 - \frac{\lambda}{3} \phi^3
\label{WZ}
\end{equation}
in conjunction with the simplest no-scale K\"ahler potential (\ref{nsK}) reproduced the Starobinsky predictions
for suitable choices of ${\hat \mu}$ and $\lambda$~\cite{ENO6}. Subsequently, many
other examples of no-scale inflationary models yielding predictions compatible with the Planck data
have been discovered and studied~\cite{ENO7}~\footnote{For a no-scale model of
Higgs inflation, see~\cite{EHX}.}. In parallel, the second data release from Planck~\cite{Planck15} has sharpened
the observational constraints on no-scale models of inflation, and interest in the observability
of tensor perturbations in the CMB has been kindled by results from BICEP2~\cite{BICEP2} and the prospects
for other experiments searching for primordial $B$-mode polarization in the CMB.

In this article we review these and other recent developments in no-scale inflation, with
particular emphasis on the r\^ole of Planck data in motivating and constraining no-scale models~\cite{FeKR,others}.
We also address the prospects for tying no-scale models of inflation more closely to string
theory and particle physics at accessible collider energies.

\section{The Effective Scalar Field Theory in No-Scale Supergravity}

An ${\cal N} = 1$ supergravity theory is characterized by its Hermitian K\"ahler 
function $K$ and holomorphic superpotential $W$ via the combination
$G \equiv K + \ln W + \ln W^*$~\cite{sugra}. The kinetic terms for scalar fields are then given in terms of the K\"ahler metric
$K_i^{j^*} \equiv \partial^2 K / \partial \phi^i \partial \phi^*_{j}$ by 
$K_i^{j^*} \partial_\mu \phi^i \partial \phi^*_j$, and (discarding $D$-terms associated with gauge interactions) the
effective potential is
\begin{equation}
V \; = \; e^G \left[ \frac{\partial G}{\partial  \phi^i} K^i_{j^*}  \frac{\partial G}{\partial  \phi^*_j} - 3 \right] \, ,
\label{effpot}
\end{equation}
where $K^i_{j^*}$ is the inverse of $K_i^{j^*}$. Equation~(\ref{effpot}) displays the major problem
for models of cosmology based on ${\cal N} = 1$ supergravity, namely that the effective potential is generically not flat, and has ``holes"
with a depth that is ${\cal O}(1)$ in natural (Planckian) units~\cite{eta}. 

It is easy to verify that the no-scale
model (\ref{nsK}) avoids this problem~\cite{GL,EENOS}, since the effective potential becomes
\begin{equation}
V \; = \; \frac{{\hat V}}{(T + T^* - |\phi|^2/3)^2} :\ {\hat V} \; \equiv \; \left| \frac{\partial W}{\partial \phi} \right|^2 \, ,
\label{effV}
\end{equation}
which is clearly positive semidefinite with no ``holes".
The kinetic terms for the scalar fields $T$ and $\phi$ in (\ref{nsK}) are
\begin{equation}
\hspace{-2cm}
{\cal L}_{KE} \; =  \; \left( \partial_\mu \phi^*, \partial_\mu T^* \right) \left(\frac{3}{(T + T^* - |\phi|^2/3)^2} \right) 
 \left( \begin{array}{cc} (T + T^*)/3 &  - \phi/3 \\ - \phi^*/3 & 1 \end{array} \right)
\left( \begin{array} {c} \partial^\mu \phi \\ \partial^\mu T \end{array} \right) \, ,
\label{no-scaleL}
\end{equation}
Assuming that the $T$ field has a
vev $\langle {\rm Re}\, T \rangle = c/2$ and $\langle {\rm Im}\, T \rangle = 0$
(we return later to a discussion how this may come about),
we may neglect the kinetic mixing between the
$T$ and $\phi$ fields in (\ref{no-scaleL}), and are left with
the following effective Lagrangian for the inflaton field $\phi$: 
\begin{equation}
{\cal L}_{\rm eff} \; = \; \frac{c}{(c - |\phi|^2/3)^2} |\partial_\mu \phi |^2 - \frac{{\hat V}}{(c - |\phi|^2/3)^2} \, ,
\label{phiL}
\end{equation}
which is the starting-point for our discussion of no-scale inflation.

\section{Starobinsky-Like Inflation from the No-Scale Wess-Zumino Model}

\begin{figure}[h!]
\centering
	\psfrag{v}[b]{\LARGE{$V/\mu^2$\phantom{www}}}
   	\psfrag{x}{\LARGE{$x$}}
   	\psfrag{a}{\large{$\lambda/\mu=1/3$}}
   	\psfrag{b}{\large{$\lambda/\mu=.33327$}}
   	\psfrag{c}{\large{$\lambda/\mu=.33330$}}
   	\psfrag{d}{\large{$\lambda/\mu=.33340$}}
   	\psfrag{e}{\large{$\lambda/\mu=.33336$}}
	\scalebox{0.6}{\includegraphics{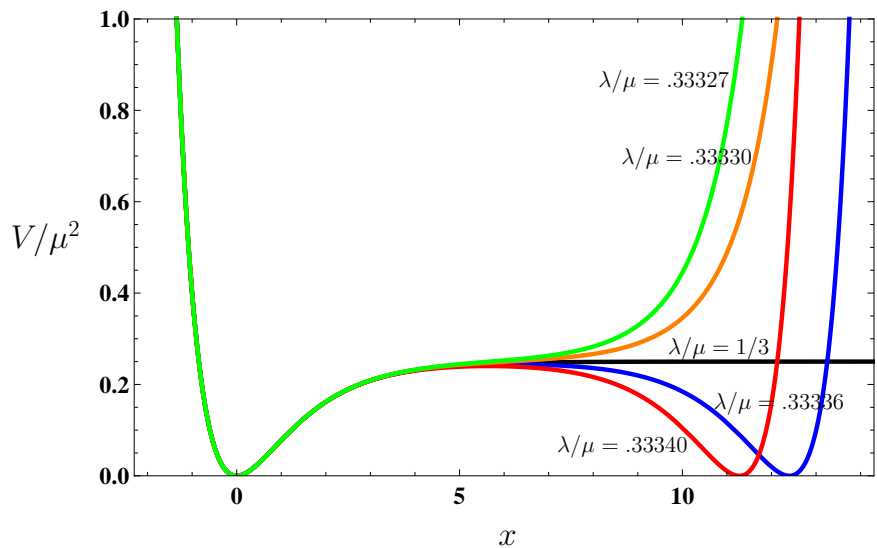}} 
\caption{\it The potential $V$ in the no-scale model described by the K\"ahler potential (\protect\ref{nsK})
and superpotential (\protect\ref{WZ}) for choices of $\lambda \sim \mu/3$ in Planck units~\protect\cite{ENO6}.}
\label{fig:ENO6}
\end{figure}

In order to study the inflaton potential in the model (\ref{nsK}, \ref{WZ}) given by (\ref{phiL}),
we first make the field transformation~\cite{ENO6}
\begin{equation}
\phi = \sqrt{3c} \tanh \left( \frac{\chi}{\sqrt{3}} \right) \, ,
\end{equation}
and introduce a rescaled parameter ${\hat \mu} = \mu\sqrt{c/3}$, in terms of which
the effective potential becomes
\begin{equation}
V = \mu^2\left |\sinh(\chi/\sqrt{3}) \left( \cosh(\chi/\sqrt{3})-\frac{3\lambda}{\mu} \sinh(\chi/\sqrt{3}) \right) \right|^2 \, .
\label{chipot}
\end{equation}
Writing $\chi$ in terms of its real and imaginary parts:
$\chi = (x + iy)/\sqrt{2}$ and choosing the specific case $\lambda = \mu/3$ (in Planck units),
we have
\begin{equation}
V (x, y)  \; = \; 
\mu^2 \frac{e^{-\sqrt{2/3}x}}{2} \sec^2(\sqrt{2/3}y)\left(\cosh{\sqrt{2/3}x})-\cos{\sqrt{2/3}y}\right) \, .
\end{equation}
At $y=0$ (fixed by the potential at large $x$), 
the potential for the real part of the inflaton takes the form 
\begin{equation}
V(x) \; = \; \mu^2 e^{-\sqrt{2/3}x} \sinh^2(x/\sqrt{6}) \, .
\label{nswzpot}
\end{equation}
This potential is displayed as the black line in Fig.~\ref{fig:ENO6}, which also shows
as coloured lines the potential
for values of $\lambda$ slightly different from the reference value $\lambda = \mu/3$ in Planck units.
The value of $\mu$ (and hence $\lambda$) is fixed by the magnitude of density perturbations
\beq
A_s = \frac{V}{24 \pi^2 \epsilon} =  \frac{ \mu^2}{8\pi^2} \sinh^4 (x/\sqrt{6}) \, ,
\eeq
where $\epsilon$ is one of the slow-roll parameters, $\simeq (1/2)(V^\prime/V)^2$. 
Using $A_s = 2.1 \times 10^{-9}$ \cite{Planck15} and $x = 5.24 - 5.45$ to insure 50-60 efolds
of inflation, we obtain $\mu = (1.9 - 2.3) \times 10^{-5}$.

Remarkably, the potential for $\lambda = \mu/3$ is {\it identical} with that in the
Starobinsky model. We recall that, starting from the modified Einstein-Hilbert action~\cite{Staro}
\begin{equation}
S=\frac{1}{2} \int d^4x \sqrt{-g} (R+R^2/6M^2) \, ,
\label{Starob}
\end{equation}
making the conformal transformation $\tilde{g}_{\mu\nu} = (1 + \varphi/3M^2) g_{\mu\nu}$
along with the field redefinition $\varphi^\prime = \sqrt{\frac{3}{2}} \ln \left( 1+ \frac{\varphi}{3 M^2} \right)$,
one obtains the action~\cite{Whitt}
\begin{equation}
S=\frac{1}{2} \int d^4x \sqrt{-\tilde{g}} \left[\tilde{R} + (\partial_\mu \varphi^\prime)^2 - \frac{3}{2} M^2 (1- e^{-\sqrt{2/3}\varphi^\prime})^2 \right] \, .
\end{equation}
This has the standard Einstein-Hilbert form with a canonically-normalized scalar field $\varphi^\prime$
whose potential is identical with (\ref{nswzpot}). Cosmological perturbations in this model
were first calculated in~\cite{MC} and, as already commented, are in perfect agreement with the Planck data~\cite{Planck13,Planck15}. For $N=50,60$, the spectral tilt is $n_s \approx 1 - 6 \epsilon + 2 \eta = 0.961 - 0.968$, where $\eta = V^{\prime\prime}/V$, and the tensor to scalar perturbation ratio is
$r \approx 16 \epsilon = 0.0030 - 0.0042$.

However, there is a price to be paid for the success of the Starobinsky model, 
namely that the measured magnitude of the scalar density perturbations
corresponds to a value of $M \ll 1$ in natural units, i.e., to an unnaturally (?) large coefficient for the $R^2$ term in (\ref{Starob}). In contrast,
in the no-scale Wess-Zumino model the magnitude of the perturbations requires $\lambda, \mu \ll 1$, which is
technically natural thanks to supersymmetry.

\section{Other Starobinsky Avatars of No-Scale Supergravity}

The Wess-Zumino superpotential is not the only way to derive
the Starobinsky model from no-scale supergravity. Indeed, it was not even the first scenario
for deriving the effective action (\ref{Starob}) from no-scale supergravity~\cite{Cecotti}, although the
connection with inflation was not made previously to~\cite{ENO6}. Ref.~\cite{Cecotti}
exhibited a no-scale supergravity example with the superpotential
\begin{equation}
W \; = \; \sqrt{3} M \phi (T - 1/2 ) \, ,
\label{Cecotti}
\end{equation}
and argued that the no-scale K\"ahler potential (\ref{nsK}) was the only consistent supergravity extension of $R + R^2$ gravity.
In the example (\ref{Cecotti}), the canonically-normalized real part $t$ of the field $T$: Re\,$T=\frac{1}{2}e^{-\sqrt{2/3}t}$
plays the r\^ole of the inflaton, and has a Starobinsky potential when $\phi$ is fixed at zero.
Many other Starobinsky-like avatars of no-scale supergravity were derived and discussed in~\cite{ENO7}.
In some of these the inflaton was identified with the matter field $\phi$ in (\ref{nsK}), and in others
it was identified with the $T$ field, which could be interpreted as a modulus of compactification~\cite{Witten}.

It is possible to represent (\ref{nsK}) in a more symmetric form~\cite{ENO7}:
\begin{equation}
K \; = \; - 3 \ln \left(1 - \frac{|y_1|^2 + |y_2|^2}{3} \right) \, ,
\label{K21symm}
\end{equation}
where the complex fields $y_{1}$ and $y_2$ are related to the fields $T, \phi$ in (\ref{nsK}) by
\begin{equation}
y_1 \; = \; \left( \frac{2 \phi}{1 + 2 T} \right) \; ; \; y_2 \; = \; \sqrt{3} \left( \frac{1 - 2 T}{1 + 2 T} \right) \, .
\label{Tphiwrite}
\end{equation}
It is important to note that the effective superpotential is modified when the coordinates are transformed as in (\ref{Tphiwrite}):
\begin{equation}
W(T, \phi) \; \to \; {\widetilde W}(y_1, y_2) \; = \; \left( 1 + {y_2}/{\sqrt{3}} \right)^3 W  \, .
\label{Wtilde}
\end{equation}
The transformation (\ref{Tphiwrite}) shows that,
at the level of the K\"ahler potential  (\ref{K21symm}) that determines the geometry
of the K\"ahler manifold, there is no real distinction
between the ``modulus" field $T$ and the ``matter" field $\phi$.
However, the distinction becomes important when one considers
the accompanying superpotential, which is an essential step in constructing a Starobinsky avatar
of no-scale supergravity.

As an example~\cite{ENO7}, if one considers the superpotential
\begin{equation}
W \; = \; M \left[ \frac{y_1^2}{2} \left(1+\frac{y_2}{\sqrt{3}} \right) - \frac{y_1^3}{3 \sqrt{3}} \right] \, ,
\label{W1}
\end{equation}
which is not obviously of Wess-Zumino form, and assumes that $\langle y_2 \rangle = 0$, one finds an effective potential
\begin{equation}
V \; = \; \frac{M^2 |y_1|^2 ~ |1 - y_1/\sqrt{3}|^2}{(1 - |y_1|^2/3)^2} \, ,
\label{V1}
\end{equation}
which yields
exactly the Starobinsky potential when rewritten in terms of the canonically-normalized inflaton field $x=\pm\sqrt{6}\tanh^{-1}(y_1/\sqrt{3})$. Moreover, transforming back to the $(T, \phi)$ basis using the inverse of
(\ref{Tphiwrite}), we find that the the K\"ahler potential and the superpotential have exactly the forms
(\ref{nsK}, \ref{WZ}). On the other hand, if one interchanges $y_1 \leftrightarrow -y_2$ and makes
the same transformation (\ref{Tphiwrite}), one finds the same K\"ahler potential (\ref{nsK}) but the
following superpotential:
\begin{equation}
W \; = \; \frac{M}{4} (T - 1/2)^2 (5 + 2 T +2 \sqrt{3} \phi ) \, .
\label{Invert}
\end{equation}
This yields the asymptotically dilatation-invariant effective potential
\begin{equation}
V \; = \; \frac{3 M^2  |T - 1/2|^2}{(T + T^*)^2}
\label{Vinvert}
\end{equation}
and, making the transformation $T = e^{\sqrt{2/3} x}/2$,
we see that this example also reproduces the Starobinsky potential,
but with the inflaton identified as the ``modulus'' field and with $\phi$ fixed at 0.

Many more examples of Starobinsky-like models with the inflaton identified as either a modulus
field or a matter field have been constructed and discussed in~\cite{ENO7,EGNO4}.
A generic issue in these and other models of inflation inspired by string theory is
how to fix the vacuum moduli, specifically in the no-scale inflationary models discussed above
the $T$ field, which may be identified with the overall compactification volume modulus.
The phenomenological approach to this problem taken in~\cite{ENO7,EGNO1} was inspired by~\cite{EKN}, namely
adding quartic terms inside the logarithm in the K\"ahler potential (\ref{nsK}):
\beq
K \; = \; - 3 \ln \left(T + T^* - \frac{|\phi|^2}{3} + \frac{(T+T^*-2c)^4 + d(T-T^*)^4}{\Lambda^2} \right) ,
\label{ekn}
\eeq
where $\Lambda$ is a mass scale assumed to be smaller than the Planck scale, and $d$ is a parameter
that breaks the invariance under the imaginary translations of the K\"ahler potential. For simplicity,
we can choose $d = 1$, in which case
the masses of the real and imaginary parts of $T$ are equal.  A non-zero mass for $T$ is most
easily obtained in this context by simply adding a constant term ${\tilde m}$
to the superpotential \cite{ENO7}, thus breaking supersymmetry. This constant term induces a gravitino mass $m_{3/2} = {\tilde m}/c^{3/2}$ and a modulus mass $m_T^2 = 288 c m_{3/2}^2/\Lambda^2$, which is hierarchically larger than the gravitino mass. For $\Lambda^2 < 0.02$, the potential for $\phi$ (or canonical $\chi$ as in (\ref{chipot}))
is indistinguishable from the Starobinsky potential.

However, stabilization terms with similar forms have not yet been derived in string theory, whereas the
corrections to (\ref{nsK}) that have been motivated by string theory, see for example~\cite{kklt}, do not stabilize the volume
modulus while maintaining the Starobinsky form of the potential necessary for inflation \cite{dudas}. 
Solving the string modulus stabilization problem lies beyond
the scope of our discussion of inflation, so we just flag it here as an important open problem.

\section{Beyond Starobinsky-Like Models in No-Scale Supergravity}

Although Starobinsky-like models are certainly highly consistent with the CMB
data from Planck and other experiments, there is scope for models with
larger values of $r$, so it is interesting to explore the scope for such a possibility
within the no-scale supergravity framework~\cite{EGNO1,EGNO2}. As an example, we consider here
a K\"ahler potential of the form~\cite{EGNO3}:
\begin{equation}
\hspace{-1.5cm}
K = - \, 3\log\left(T+\bar{T} - \frac{\left[\cos\theta(T+\bar{T}-1)-\sin\theta(T-\bar{T})^2\right]^2}{\Lambda^2} \right)+\frac{|\phi|^2}{(T+\bar{T})^3} \, ,
\label{ourK}
\end{equation}
where the second term is typical of how a matter field with modular weight $w = 3$
would appear in an orbifold compactification of string~\cite{casas}, and we postulate a
superpotential of the form
\begin{equation}
W \; = \; \sqrt{\frac{3}{4}}\,\frac{m}{a}\phi(T-a) \, ,
\label{ourW}
\end{equation}
where $a$ is some coefficient $\leq1$. In this case, we have a model where the modulus, $T$
plays the role of the inflaton, and one linear combination of the real and imaginary parts of $T$,
defined by the angle $\theta$,  is
stabilized for $\Lambda < 1$. 
It is easy to check that $\phi$ is driven to zero,
for which value the effective potential takes the simple form
\begin{equation}
V = \frac{3 m^2}{4 a^2} |T-a|^2 \, .
\end{equation}
We decompose $T$ into its real and imaginary parts $(\rho, \sigma)$,
where $\rho$ is normalized canonically and $\sigma$ is canonical at the minimum
when $\rho = 0$:
\begin{equation}
T \; = \; a\left(e^{-\sqrt{\frac{2}{3}}\rho}+i\sqrt{\frac{2}{3}}\,\sigma\right) \, .
\label{rhosigma}
\end{equation}
The potential is minimized when $T=a$, in which case
the effective Lagrangian is given by
\begin{equation}
\mathcal{L} \; = \; \frac{1}{2}\partial_{\mu}\rho\partial^{\mu}\rho+\frac{1}{2}e^{2\sqrt{\frac{2}{3}}\rho}\partial_{\mu}\sigma\partial^{\mu}\sigma - 
\frac{3}{4}m^2\left(1-e^{-\sqrt{\frac{2}{3}}\rho} \right)^2 - \frac{1}{2}m^2\sigma^2 \, ,
\label{effL}
\end{equation}
and it is easy to see that the minimum of this effective potential is at
\begin{equation}
\rho_0 \; = \; \sigma_0 \; = \; 0 \, .
\label{minimum}
\end{equation}
When $\rho$ is at its minimum, the effective Lagrangian for $\sigma$ is
\begin{equation}
\mathcal{L}=\frac{1}{2}\partial_{\mu}\sigma\partial^{\mu}\sigma -  \frac{1}{2}m^2\sigma^2 \, ,
\label{Lsigma}
\end{equation}
and we recover the minimal chaotic inflationary model with a quadratic potential \cite{FeKR,EGNO1,EGNO2}. Conversely, 
when $\sigma$ is at its minimum, the effective Lagrangian for $\rho$ is
\begin{equation}
\mathcal{L} \; = \; \frac{1}{2}\partial_{\mu}\rho\partial^{\mu}\rho - 
\frac{3}{4}m^2\left(1-e^{-\sqrt{\frac{2}{3}}\rho} \right)^2 \, ,
\label{Lrho}
\end{equation}
which yields the familiar Starobinsky potential~\cite{Staro}.

\begin{figure}[h!]
\centering
	\scalebox{0.8}{\includegraphics{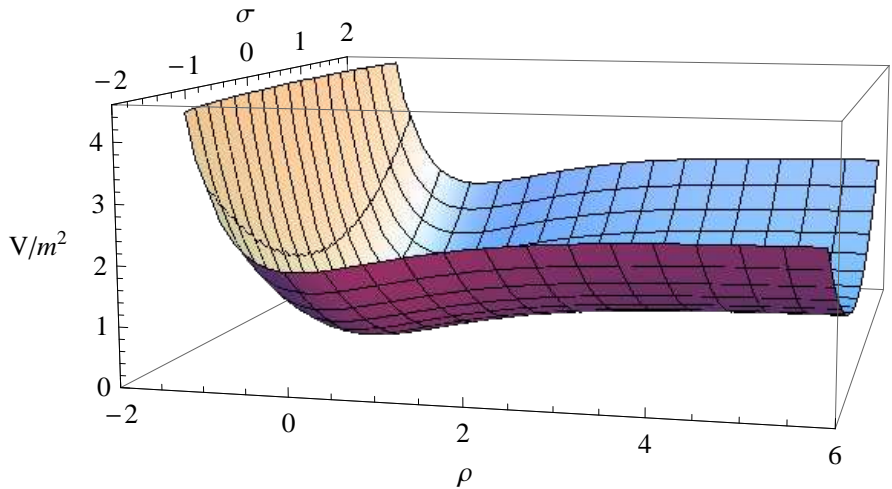}} 
		\caption{\it The effective potential of the no-scale model (\protect\ref{effL}),
		displaying its Starobinsky form in the $\rho$ direction and its quadratic form in the $\sigma$ direction~\protect\cite{EGNO2}.}
		\label{fig:EGNO2-1}
\end{figure}

The potential (\ref{effL}) for the choice $a=1/2$
is shown in Fig.~\ref{fig:EGNO2-1}, where we see the 
Starobinsky form in the $\rho$ direction (\ref{Lrho}) and the quadratic form in the
$\sigma$ direction (\ref{Lsigma})~\footnote{A more complete discussion of this potential, including
its behaviour as a function of $\phi$, is given in~\cite{EGNO2}.}.

This model has two dynamical fields coupled through the kinetic term of the Largrangian (\ref{effL}), and a full discussion of their behaviour during inflation requires a
complete two-field analysis~\cite{2field,Turzynski}, which we summarize in the next Section.
For now, we just note that for small $\Lambda$ the $\theta$-dependent stabilization terms in (\ref{ourK})
reduce the model to a family of nearly single-field models 
characterized by an angle $\theta$ in the $({\rm Re}\,T, {\rm Im}\,T)$ plane.
If the coefficient $\Lambda^{-1}$ of the quartic stabilization term is large enough, the inflaton
trajectory is confined to a narrow valley in field space, like a bobsleigh running down
a narrow track.

\begin{figure}[h!]
\centering\
	\psfrag{n}[b]{\LARGE{$n$}}
	\psfrag{s}{$s$}
	\psfrag{r}[b]{\LARGE{$r$}}
	\scalebox{0.7}{\includegraphics{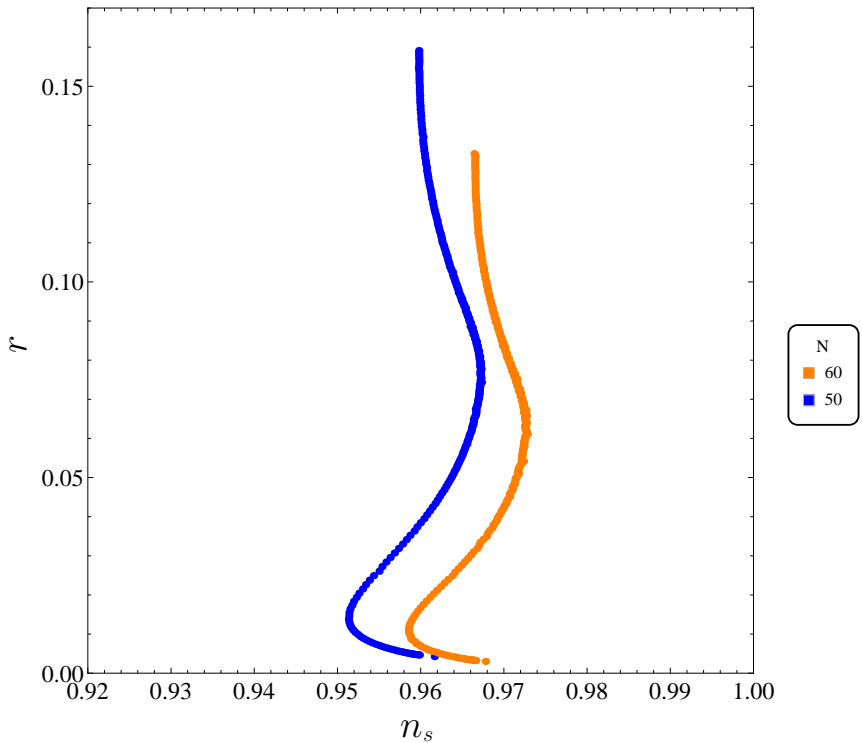}} 
	\caption{\it The parametric curve of predictions $(n_s(\theta),r(\theta))$ in the no-scale inflationary model
	(\protect\ref{effL}), assuming that the angle $\theta$ in the complex $T$ field is constrained by the
	stabilization terms in (\protect\ref{ourK}) with $\Lambda = 0.1$~\protect\cite{EGNO3}. The values $\theta = 0, \pi/2$
	correspond to the tops (bottoms) of the curves for $N = 50, 60$ e-folds after horizon exit (blue and yellow, respectively).} 
	\label{fig:EGNO2-2}
\end{figure}

It is clear that $n_s$ and $r$ must depend on the initial condition for the complex inflaton field $T$,
and particularly the value of $\theta$. As an example, we consider initial conditions in the $(\rho,\sigma)$ plane
that lead to $N+10$ e-foldings of inflation, for $N=50,60$, assuming $\phi = 0$ and setting $\Lambda = 0.1$. 
The resulting $\theta$ dependences of $r$ and $n_s$ are shown in
Fig.~\ref{fig:EGNO2-2}.
Here we see clearly how the model (\ref{effL}) interpolates between the
limits of quadratic and Starobinsky-like inflation as $\theta$ increases from $0 \to \pi/2$~\cite{EGNO3}.
The Planck 2015 data~\cite{Planck15} disfavour the small values of $\theta$ that yield $r \ga 0.1$.

Values of $r$ that are smaller than in the Starobinsky model are possible in other models related to no-scale
supergravity~\cite{ENO7}. The Starobinsky potential can be expressed parametrically
\begin{equation}
V \; = \; A \left( 1 - e^{-Bx} \right)^2 \, ,
\label{StaroAB}
\end{equation}
where $x$ is canonically normalized, the value of $A$ is determined by the magnitude of the scalar density
perturbations, and $B = \sqrt{2/3}$ in the Starobinsky model. The inflationary predictions are
derived at large $x$ where the potential is dominated by the constant and leading term $\propto e^{-Bx}$ in (\ref{StaroAB}). One can consider phenomenological generalizations of (\ref{StaroAB}) where
\begin{equation}
V \; = \; A \left( 1 - \delta e^{-Bx} + {\cal O}(e^{-2Bx}) \right) \, ,
\label{StaroAlambdaB}
\end{equation}
and $\delta$ and $B$ treated as free parameters that  may differ from the
Starobinsky values $\delta = 2$ and $B = \sqrt{2/3}$. At leading order in $e^{-Bx}$ one finds
\begin{equation}
n_s \; = \; 1 - 2 B^2 \delta e^{-Bx}, \; r \; = \; 8 B^2 \delta^2 e^{-2Bx}, \; N_* \; = \; \frac{1}{B^2 \delta} e^{+ Bx} \, .
\label{predictions}
\end{equation}
implying
\begin{equation}
n_s \; = \; 1 - \frac{2}{N_*} \; ,\quad  r \; = \; \frac{8}{B^2 N_*^2} \, .
\label{relations}
\end{equation}
These predictions are independent of $\delta$, and the prediction for $n_s$ is independent of
$B$. The only model-dependence is that $r$ depends on $B$.

Within the no-scale framework, different values of $B$ could
be obtained in models with multiple moduli $T_i$ that share the no-scale property~\cite{ENO7} 
\begin{equation}
K \; \ni \; - \sum_i \, N_i \, \ln (T_i + T_i^*) : \; \; N_i > 0, \; \; \sum_i \, N_i = 3 \, . 
\label{multimoduli}
\end{equation}
If one identifies the inflaton with one of the moduli $T_i$, one finds that has a potential of
the form (\ref{StaroAB}) with
\begin{equation}
B \; = \; \sqrt{\left(\frac{2}{N_i}\right)} \, ,
\label{generalB}
\end{equation}
and hence
\begin{equation}
r \; = \; 
\frac{4 N_i}{N_*^2} \, .
\label{generalr}
\end{equation}
The leading alternative to the single-modulus case with $N_i = 3$
may be that with three moduli $T_i$, each with $N_i = 1$, one of which is identified with the inflaton.
In this case $r$ would be a factor of 3 smaller than in the
Starobinsky model. This example shows that, within the class of no-scale models discussed here,
an eventual measurement of $r$ might provide some observational clues
to the form of string compactification.

\section{Two-Field Effects}

Since the building-blocks of supersymmetric models are complex scalar fields, in general,
supersymmetric models of inflation must take multi-field effects into account~\cite{2field}. For example,
the original no-scale model (\ref{nsK}) has four field components in general, as does the model
(\ref{ourK}) discussed in the previous Section. Early studies of no-scale models took the (over-)simplified
approach of fixing some (combination) of the field components, as was done in (\ref{ourK}).
What happens if one relaxes this assumption, and considers the full multi-field dynamics
of the inflaton field?

\begin{figure}[!h]
\centering
	\hspace{-15pt}
	\psfrag{a}{$\sigma$}
		\scalebox{0.45}{\includegraphics{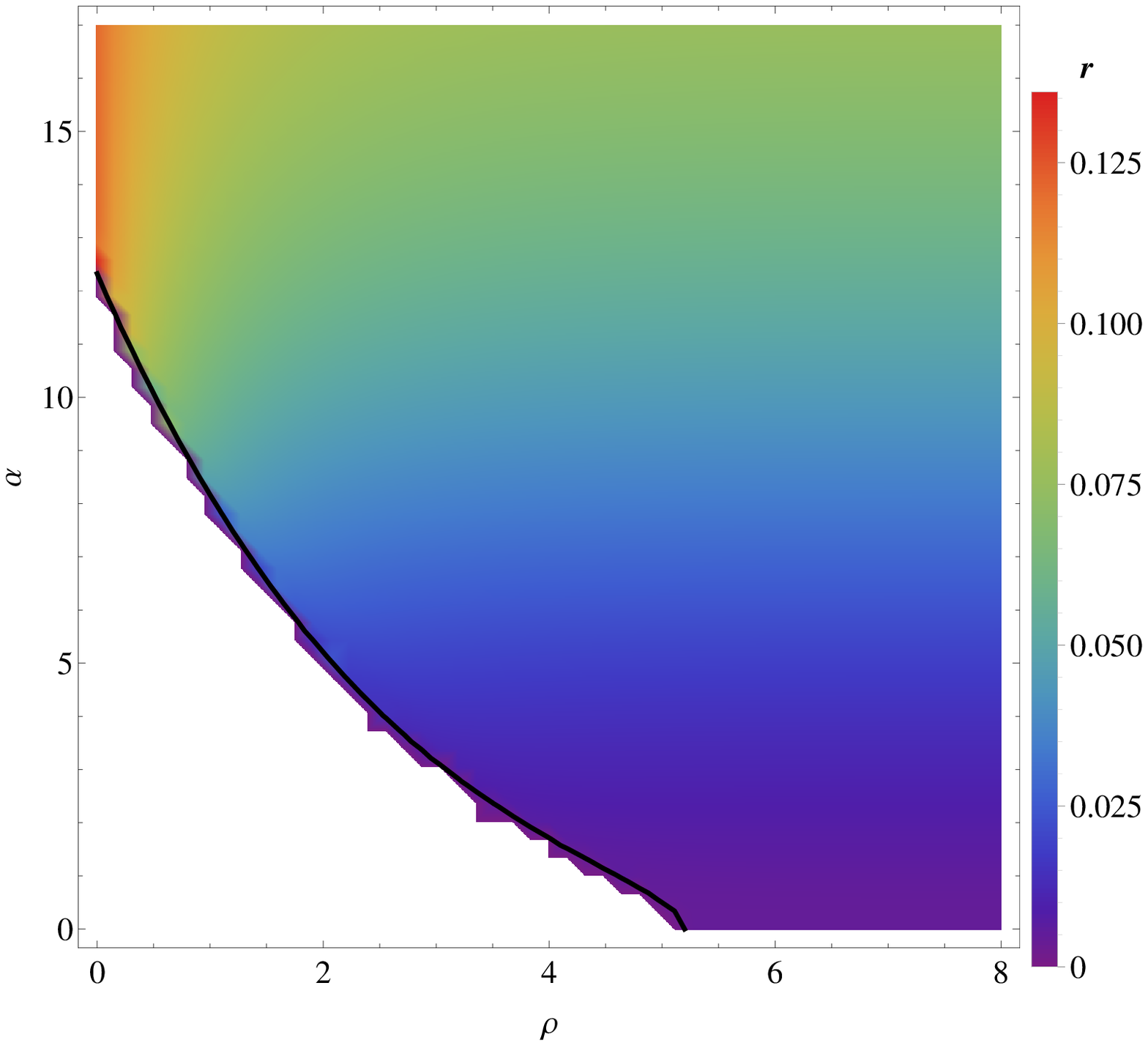}}
	\scalebox{0.45}{\includegraphics{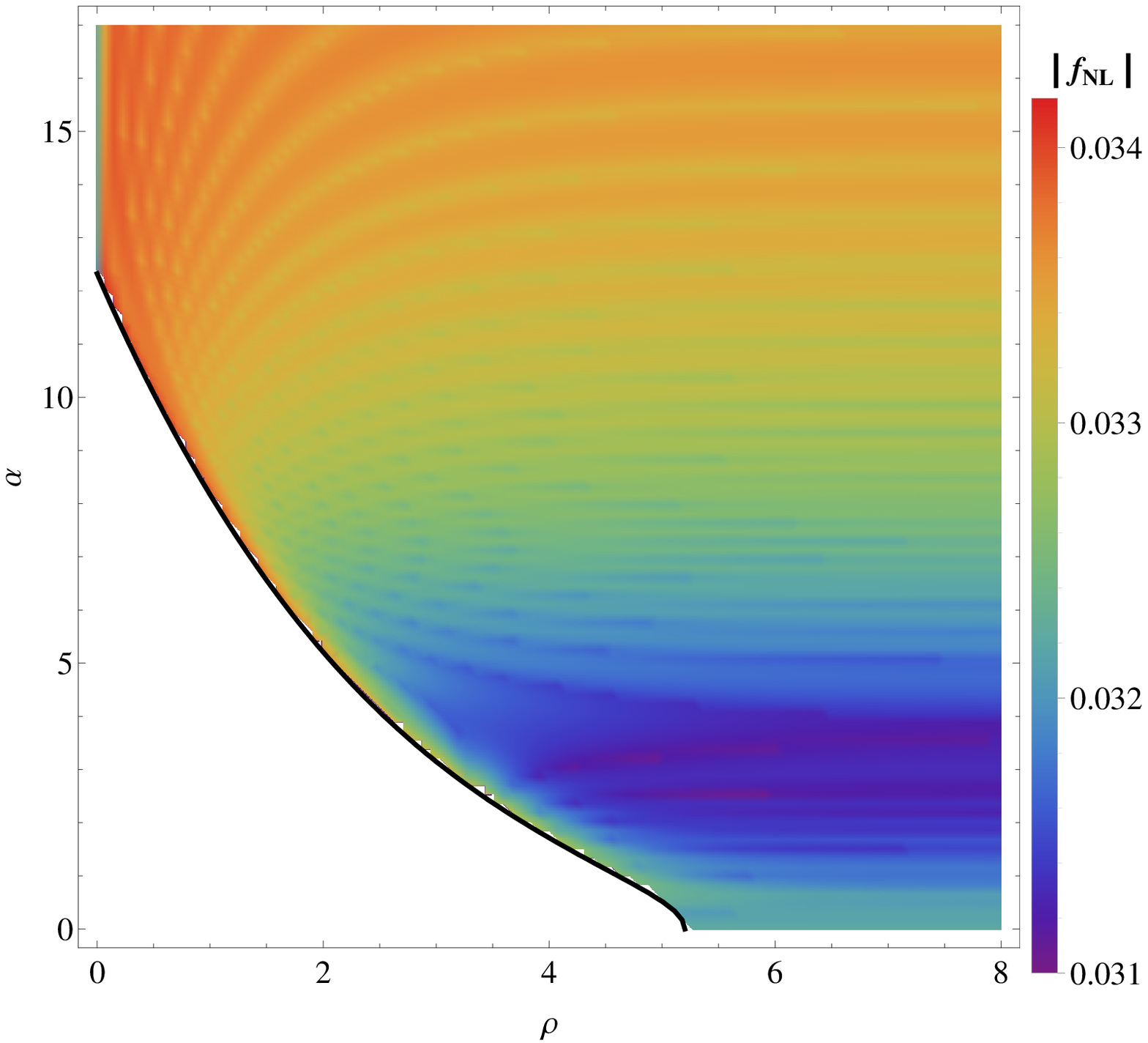}}
	\caption{\it Left panel: The tensor-to-scalar ratio $r$ from the two-field analysis in the $(\rho,\sigma)$ plane,
	assuming $N = 50$ and $\Lambda^2 = 10$. Right panel: The corresponding values of the non-Gaussianity 
	measure $f_{\rm NL}$~\protect\cite{EGNO3}.}
	\label{fig:EGNO3}
\end{figure}

In general, it is known that the inflaton field trajectory will be curved and that, as a result,
isocurvature fluctuations perpendicular to the direction of inflaton motion motion
source adiabatic perturbations as the field trajectory evolves towards the global minimum.
This extra source of adiabatic scalar perturbations tends to suppress the tensor-to-scalar ratio $r$,
and may in addition source non-Gaussian effects such as $f_{\rm NL}$~\cite{2field}. We have studied such
effects in the specific two-field no-scale model introduced in the previous Section~\cite{EGNO3}, and some
representative results are shown in Fig.~\ref{fig:EGNO3}~\footnote{Two-field effects were already 
incorporated in Fig.~\ref{fig:EGNO2-2} in the case where the
stabilization term in (\ref{ourK}) has the value $\Lambda = 0.1$, i.e., with strong stabilization
of the inflaton trajectory, although they were unimportant  in this case.}.
The left panel of Fig.~\ref{fig:EGNO3} shows predictions for the tensor-to-scalar ratio $r$
as a function of the starting-point in the $(\rho, \sigma)$ plane for $\Lambda^2 = 10$, i.e., with the
stabilization term in (\ref{ourK}) switched off (almost). We see in the left panel that,
as expected, $r$ is reduced compared to what might have been expected from a naive
single-field analysis. Only very close to the vertical axis ($\theta \sim 0$) does $r$ approach
the value predicted by the chaotic inflationary model with a quadratic potential. In most
of the plane of possible initial conditions, $r$ takes values comparable to those in the
Starobinsky model. The right panel shows that these Planck-compatible
values of $r$ are not accompanied by large non-Gaussianity~\cite{EGNO3}: in all the plane, $|f_{\rm NL}| \sim 0.03$,
well within the Planck bound on this measure of non-Gaussianity: $f_{\rm NL} = 0.8 \pm 5.0$~\cite{Ade:2015ava}.

\section{Inflaton Decays}

As is well-known, the predictions of any specific model of inflation depend
on $N_*$, the number of e-folds at some reference scale $k_*$. For example, in
Starobinsky-like models one has
\begin{equation}
n_s \; \simeq \; 1 - \frac{2}{N_*}, \; \; r \; \simeq \; \frac{12}{N_*^2} \, .
\label{N*dependences}
\end{equation}
The small values of $r$ in (\ref{N*dependences}) are unlikely to be probed in the near future,
but the Planck measurement of $n_s$ is already imposing interesting constraints on $N_*$~\cite{EGNO5}.
This is related to inflationary model parameters via~\cite{reheat}
\begin{equation}
\hspace{-2.2cm}
N_* \; = \; 66.9 - \ln\left(\frac{k_*}{H_0 a_0}\right) + \frac{1}{4}\ln\left(\frac{V_*^2}{M_P^4\rho_{\rm end}}\right)
 + \frac{1-3w_{\rm int}}{12(1+w_{\rm int})}\ln\left(\frac{\rho_{\rm reh}}{\rho_{\rm end}}\right) - \frac{1}{12}\ln g_{\rm reh}\ ,
\label{howmany}
\end{equation}
where we have ignored the possibility of entropy generation after reheating~\cite{DM}, $H_0$ and $a_0$ are the present 
Hubble expansion rate and cosmological scale factor, respectively, 
$V_*$ is the inflationary energy density at the reference scale, 
$\rho_{\rm end}$ and $\rho_{\rm reh}$ are the energy densities at the end of inflation and after reheating, 
$w_{\rm int}$ is the {\it e-fold} average of the equation-of-state parameter $w$ during the thermalization epoch, 
and $g_{\rm reh}$ is the number of equivalent bosonic degrees of freedom after reheating:
$\rho_{\rm reh} \equiv (\pi^2/30)g_{\rm reh}T_{\rm reh}^4$.
The values of the reheating energy density $\rho_{\rm reh}$ and temperature $T_{\rm reh}$ depend, in turn,
on the rate of inflaton decay, $\Gamma_\phi$, and we find~\cite{EGNO5} that
\begin{equation}
\label{rhoreh}
\rho_{\rm reh} \; \simeq \; \frac{4}{3}(0.655-1.082\ln\delta)^{-2}(1+w_{\rm eff})^{-2}M_P^2\Gamma_{\phi}^2 \, ,
\end{equation}
where $\delta$ parametrizes the approach to complete thermalization:
\begin{equation}
\label{omegareh}
\delta \; \equiv \; 1 - \frac{\rho_{\gamma}}{\rho_{\phi}+\rho_{\gamma}} \, ,
\end{equation}
and $w_{\rm eff}$ is the {\it time} average of the equation-of-state parameter $w$ during the thermalization epoch, 
which we find to be $\simeq 0.271$ for $\Gamma_\phi/m \ll 1$.
As a result, $N_*$ depends on the inflaton decay rate as follows:
\begin{equation}
\hspace{-2cm}
N_* \; \ni \;  \frac{1-3w_{\rm int}}{6(1+w_{\rm int})}[ \ln\left(\Gamma_{\phi}/m\right) \, - \, \ln(1 + w_{\rm eff}) \, - \, 2 \ln(0.655 - 1.082 \ln \delta) ]  \, + \, \dots \, , \,
\end{equation}
where the $\dots$ represent other terms in the full expression for $N_*$ given in~\cite{EGNO5},
where it is also shown that $w_{\rm int} \simeq 0.782/\ln(2.096m/\Gamma_{\phi})$ in
Starobinsky-like models.

\begin{figure}[h!]
\vspace{0.3cm}
\centering
\hspace{0.5cm}
	\scalebox{0.65}{\includegraphics{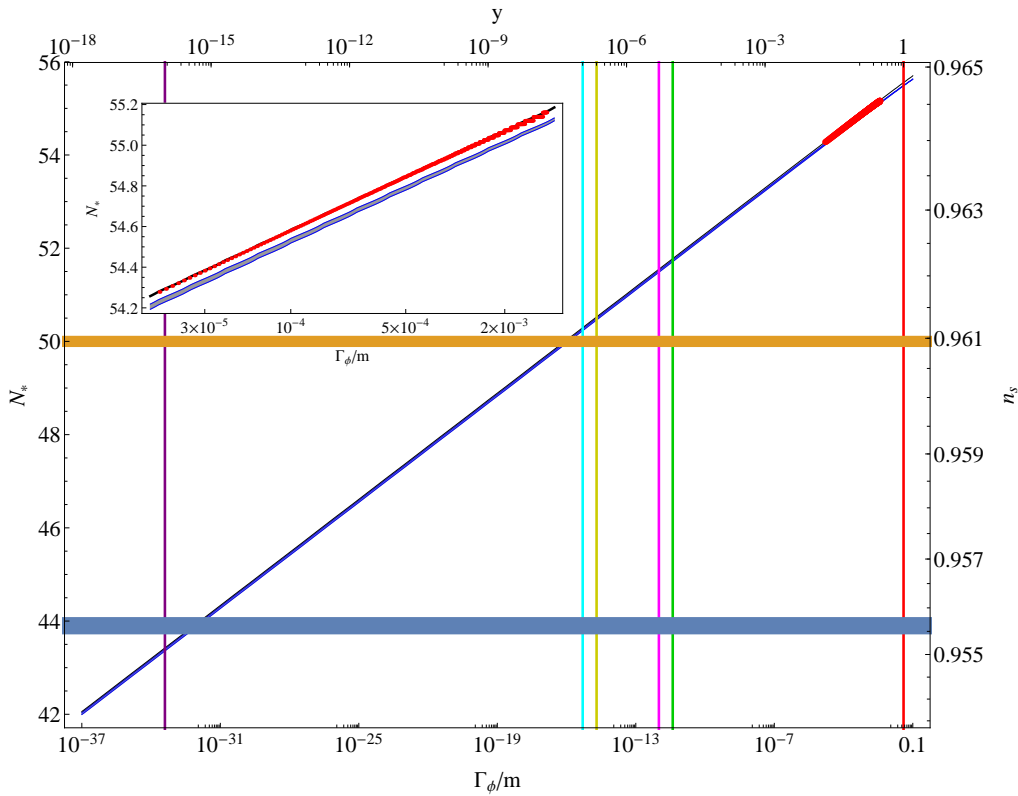}}
	\caption{\it The values of $N_*$ in no-scale Starobinsky-like models (left axis) and $n_s$ (right axis)
	for a wide range of decay rates $\Gamma_\phi/m$ (bottom axis) and corresponding two-body couplings $y$ (top axis). The
	diagonal red line segment shows full numerical results for $\delta=0.002$ over a restricted range of $\Gamma_\phi/m$, 
	which are shown in more detail in the insert, and the diagonal blue strip represents an analytical approximation 
	for $10^{-3}<\delta<10^{-1}$. The difference between these results is indistinguishable in the main plot, 
	but is visible in the insert. The vertical
	coloured lines correspond to specific models of inflaton decay discussed in~\protect\cite{EGNO5}, 
	and the horizontal yellow (blue) lines show the
	68 and 95\% CL lower limits from the Planck 2015 data~\protect\cite{Planck15}, which differ slightly in other no-scale models,
	as also discussed in~\protect\cite{EGNO5}.} 
	\label{fig:EGNO5-1}
\end{figure}

Fig.~\ref{fig:EGNO5-1} shows the corresponding numerical relation between $\Gamma_\phi$, $N_*$ and $n_s$
in Starobinsky-like models including those based on no-scale supergravity. The diagonal red and blue lines are numerical
results and analytic approximations, respectively, which agree quite well, and the vertical lines represent specific
models of inflaton decay discussed in~\cite{EGNO5}. The upper axis shows values of the two-body
decay coupling $y$ corresponding to $\Gamma_\phi$ via the relation $\Gamma_{\phi} = m|y|^2/8\pi$,
for values of the coupling $y$ ranging from $y = 1$ (vertical red line) to the value $y \simeq10^{-16}$
(vertical purple line), which would correspond to a reheating temperature $T_{\rm reh} \simeq 10$~MeV,
below which the successful conventional Big Bang nucleosynthesis calculations would need to be
modified substantially. In order to avoid the overproduction of gravitinos, whose decays could also aversely
affect big bang nucleosynthesis and overpopulate the Universe with dark matter particles \cite{bbb},
one may require $y < 10^{-5}$, corresponding to the vertical green line in Fig.~\ref{fig:EGNO5-1}.
The vertical magenta line corresponds to the rate for three-body decays involving top quarks that dominate
in some no-scale models~\cite{EGNO4}, and the vertical yellow and magenta lines bracket the expected range for decays
into gauginos that may be found in models with non-trivial gauge kinetic functions $f_{\alpha \beta} \ne 0$.

\begin{figure}[h!]
\centering
\scalebox{0.6}{\includegraphics{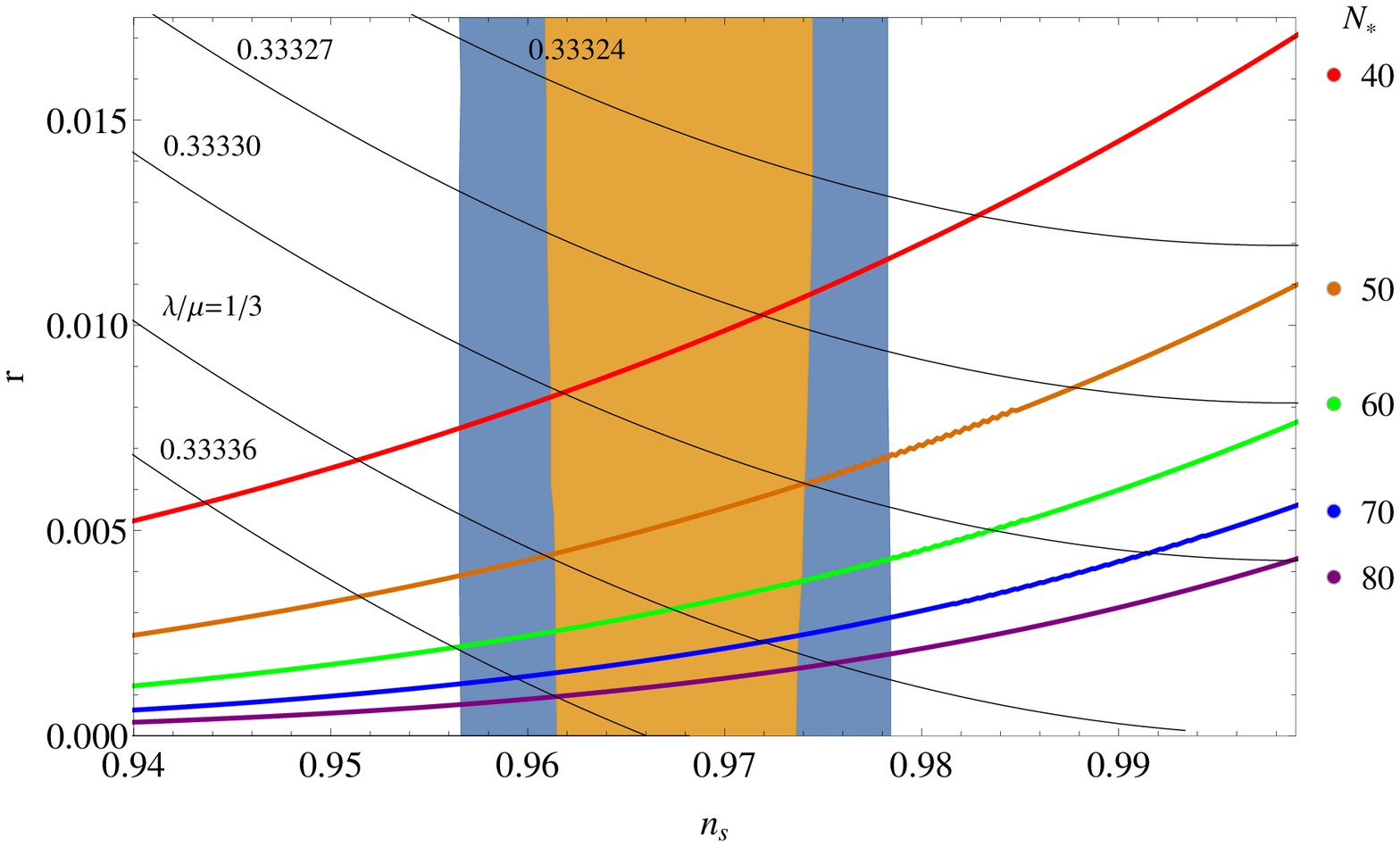}}\\
\hspace{-0.98cm}
\scalebox{0.62}{\includegraphics{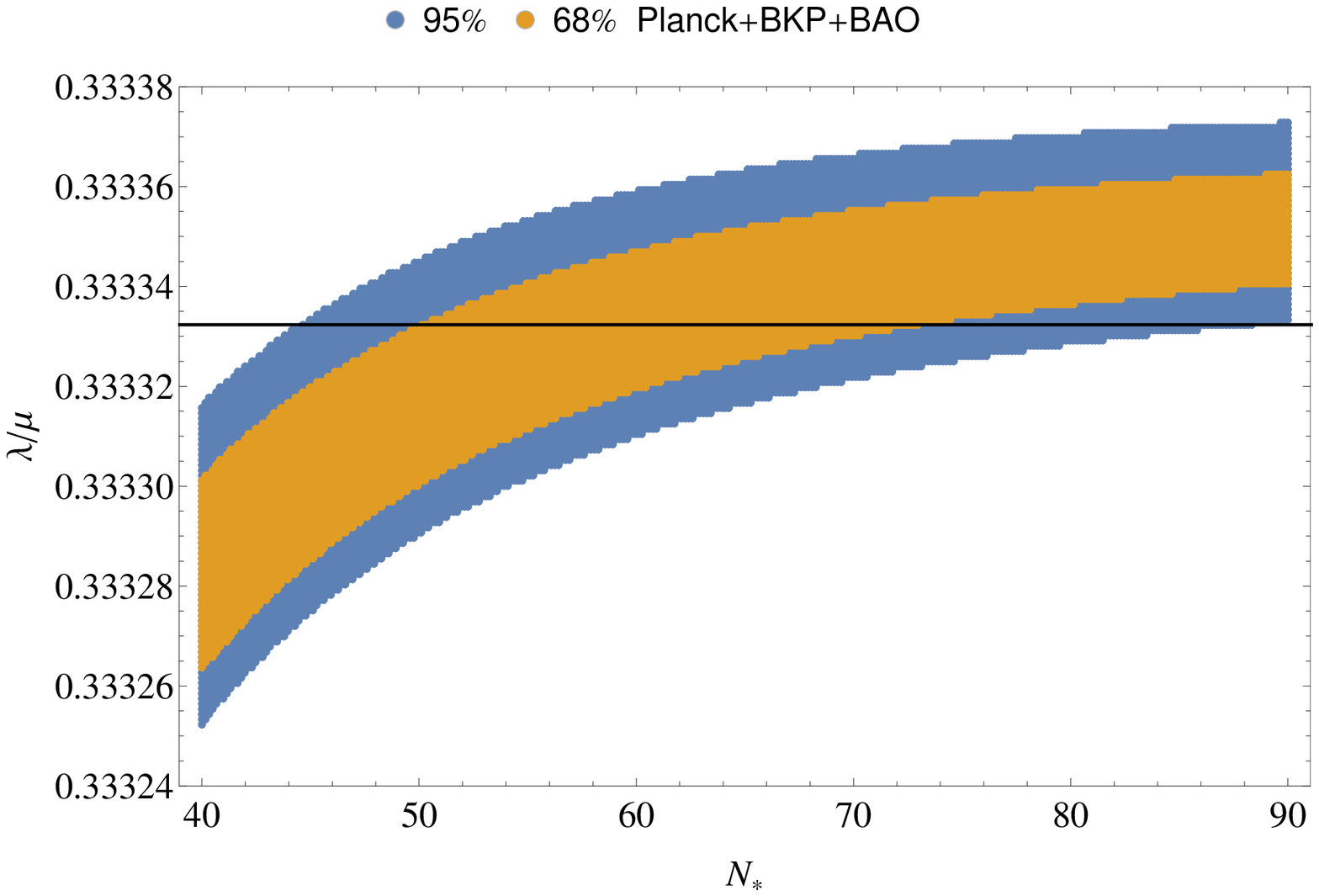}}
	\caption{\it The Planck 2015~\protect\cite{Planck15}
	68\% and 95\% CL regions (yellow and blue, respectively) in the $(n_s, r)$ plane (upper panel)
	and the $(N_*, \lambda/\mu)$ plane (lower panel)
	for the no-scale inflationary model with the Wess-Zumino superpotential
	(\protect\ref{WZ})~\protect\cite{EGNO5}. The black lines in the upper panel are contours of $\lambda/\mu$,
	and the coloured lines are contours of $N_*$.
	The horizontal black line in the lower panel is for $\lambda/\mu = 1/3$, the value that reproduces the
	inflationary predictions of the Starobinsky model~\protect\cite{ENO6}.} 
	\label{fig:EGNO5-2}
\end{figure}

Fig.~\ref{fig:EGNO5-2} confronts the Planck 2015 constraints~\cite{Planck15} - the yellow shaded regions are favoured at the
68\% CL and the blue shaded regions are allowed at the 95\% CL - with the predictions of the simplest
Wess-Zumino no-scale model~\cite{ENO6} for different values of $N_*$ and $\lambda/\mu$ near the value 1/3 that
reproduces Starobinsky-like predictions~\cite{EGNO5}. We see in the upper panel that the Planck 2015 constraints are almost
independent of $r$ in the displayed range of $r$, and that the contours of fixed $N_*$ (coloured lines)
intersect the contours of fixed $\lambda/\mu$ (black lines) at acute angles. Consequently, as seen in the
lower panel, the Planck 2015 constraints on $N_*$ depend quite strongly on the value of $\lambda/\mu$.
For the Starobinsky value $\lambda/\mu = 1/3$ we find lower bounds
\begin{equation}
N_* \; \sim \; 50 \; \; (68\%~CL), \; \; \sim 44 \; \; (95\%~CL) \, ,
\end{equation}
which are shown as horizontal yellow and blue lines, respectively, in Fig.~\ref{fig:EGNO5-1}.
Their intersections with the diagonal blue line give the corresponding lower bounds on $\Gamma_\phi/m$ and $y$.
We see that the lower bounds are already stronger than those imposed by successful Big Bang nucleosynthesis
(vertical purple line) and the 68\% CL lower limit approaches the upper limit on $y \sim 10^{-5}$ suggested by
gravitino decays (vertical green line). Thus, the Planck data~\cite{Planck15}
already impose interesting constraints on inflaton decays in no-scale
models: other examples are discussed in~\cite{EGNO5}.

\section{From the LHC to String via No-Scale Inflation}

One of the biggest challenges in constructing models of inflation is to relate them to particle
physics at accessible (collider) energies. One connection may be provided in principle by the
discussion of inflaton decays in the previous section~\cite{EGNO4}. For example, in models where the
inflaton is identified with the supersymmetric partner of a singlet (right-handed) neutrino
in a see-saw model of neutrino masses, the constraints on the two-body decay coupling
$y$ have potential implications for low-energy observables such as flavour-changing
lepton transitions as well as scenarios for neutrino mixing. Another possible connection
arises via supersymmetry breaking. For example, if the gauge kinetic function $f_{\alpha \beta}$
depends same non-trivially on the inflaton field $\phi$:
\begin{equation}
d_{g,\phi} \equiv \langle {\rm Re}\,f\rangle^{-1}\left|\left\langle\frac{\partial f}{\partial \phi}\right\rangle\right| \ne 0 \, ,
\label{dgphi}
\end{equation}
the inflaton decays into Standard Model gauge bosons and gauginos are given by~\cite{EGNO4}
\begin{equation}
\Gamma(\phi\rightarrow gg) = \Gamma(\phi \rightarrow \tilde{g}\tilde{g}) = \frac{3d_{g,\phi}^2}{32\pi}\frac{m^3}{M_P^2}\,,
\label{gaugedecay}
\end{equation}
and the gaugino masses are given by
\begin{equation}
m_{1/2} = {\cal O}(1) \times d_{g,\phi} \times m_{3/2} \, .
\label{m12m32}
\end{equation}
As already commented, the vertical yellow and magenta lines in Fig.~\ref{fig:EGNO5-1}
bracket the range of $y$ and hence $\Gamma_\phi/m$ found in Starobinsky-like
no-scale models, which correspond to $N_* \in (50.5, 51.5)$, with the upper end of
this range corresponding to $d_{g,\phi} = {\cal O}(1)$ and hence $m_{1/2} = {\cal O}(m_{3/2})$.

Additionally, different assignments for the inflaton and matter fields lead to different
possibilities for the pattern of supersymmetric particle masses, via their
dependences on the model-dependent soft supersymmetry-breaking scalar masses $m_0$, gaugino masses $m_{1/2}$ and
bilinear and trilinear scalar couplings $B_0$ and $A_0$~\cite{EGNO4}. Among the possibilities are
the original {\it no-scale} boundary conditions $m_0 = B_0 = A_0 =0$ \cite{LN,EKNN}, {\it CMSSM}-like
boundary conditions in which $m_0, B_0$ and $A_0$ are non-zero and universal for
different scalar species and determined by the gravitino mass \cite{bfs}, {\it mSUGRA}-like boundary conditions in which $m_0 + B_0 = A_0$ \cite{bfs,vcmssm}, etc.
Thus, if supersymmetry is discovered and sparticle masses measured at accelerators, it may
be possible constrain models of inflation, and vice versa if models of inflation can be constrained.

\begin{figure}[h!]
\centering
\vskip -0.9in
\includegraphics[scale=0.6]{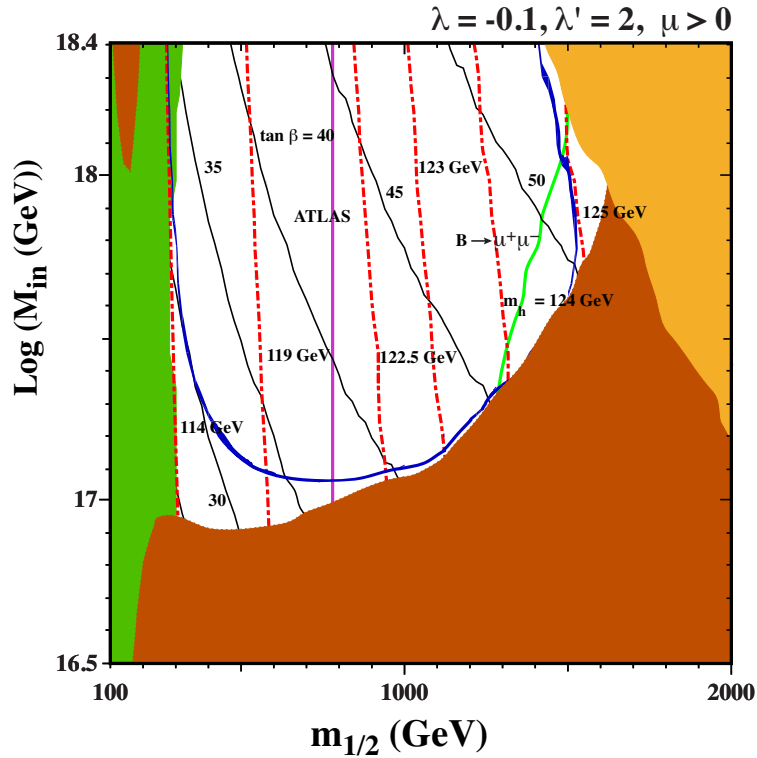}
\caption{\it The $(m_{1/2}, M_{in})$ plane in the SU(5) no-scale
model with $m_0 = A_0 = B_0=0$ at $M_{in}$, for $\lambda^\prime = 2$ and $ \lambda = - 0.1$~\protect\cite{ENO8}.
The brown shaded regions are excluded because the LSP is charged,
the green shaded regions are excluded by $b \to s \gamma$, the renormalization-group equations are
unstable in the ochre shaded region of the upper panel. The relic cold dark matter density lies within the cosmological range
along the narrow dark blue strips. The near-vertical continuous purple line is the ATLAS 95\% CL lower limit on
the gaugino mass $m_{1/2}$, the dash-dotted (red) lines are contours of $m_H$ 
as calculated using {\tt FeynHiggs~2.10.0}~\protect\cite{FH},
and the solid diagonal green lines marks a 95\% CL upper limit on BR($B_s \to \mu^+ \mu^-$).
The continuous (black) lines in the upper panel are contours of $\tan \beta$,
the ratio of supersymmetric Higgs vevs~\protect\cite{ENO8}.
}
\label{fig:ENO8}
\end{figure}

An example is given in Fig.~\ref{fig:ENO8}, which displays results for a no-scale scenario
with $m_0 = B_0 = A_0 = 0$ and $m_{1/2} \ne 0$ at some input renormalization scale $M_{in}$
in an SU(5) GUT model with superpotential terms $W \ni \lambda {H} {\Sigma} {\bar H} + (\lambda^\prime/6) {\rm Tr}{\Sigma^3}$, where
$H, {\bar H}$ and $\Sigma$ are $\mathbf{5}, \mathbf{\bar 5}$ and $\mathbf{24}$
Higgs representations, respectively, for the representative values $\lambda = - 0.1, \lambda^\prime = 2$
discussed in~\cite{emo2,ENO8}. We see that, within this particular model, only a restricted range of
$m_{1/2} \in (800, 1500)$~GeV is consistent with the LHC data. The relations
(\ref{m12m32}, \ref{gaugedecay}, \ref{dgphi}) show how this type of constraint can then be
applied to models of inflation and string compactification.

\section{Summary and Prospects}

As we have reviewed in this article, no-scale supergravity is an attractive  framework
for constructing models of inflation, since supersymmetry accommodates naturally the
required hierarchy between the scale of inflation and the Planck scale~\cite{Cries}, gravity must be
incorporated in any discussion of cosmology, and no-scale supergravity~\cite{CFKN} emerges from
generic string compactifications~\cite{Witten} and yields an effective potential that is positive
semi-definite~\cite{GL,EENOS}. We have discussed various no-scale inflationary scenarios, reviewing
how they naturally yield an effective potential, and hence predictions for $n_s$ and $r$, that are
coincident~\cite{ENO6,ENO7} with the Planck-friendly Starobinsky model based on $R + R^2$ gravity~\cite{Staro}.
Moreover, no-scale models achieve this in a technically natural way, via small superpotential couplings
rather than surprisingly large non-minimal gravitational couplings as in the Starobinsky and Higgs
inflation models.

The no-scale framework is, moreover, more flexible, being able to accommodate models
intermediate between the Starobinsky model and a quadratic potential suitable for chaotic inflation~\cite{EGNO1,EGNO2},
which could yield larger values of $r$ that are still compatible with the constraints
from Planck and other experiments. As in other supersymmetric models of inflation,
it is necessary to take into account two-field effects: in the no-scale models we study~\cite{EGNO3},
we find that these tend to reduce $r$ without
generating large non-Gaussianity.

In principle, no-scale inflation could provide a phenomenological bridge between
string theory and collider physics~\cite{EGNO4}. In addition to the above-mentioned model-dependence of $r$,
the value of $n_s$ is related directly in Starobinsky-like models to the
number of e-folds during inflation~\cite{EGNO5}, which is in turn sensitive to the rate of inflaton decay
and thereby the assignment of the inflaton as a modulus or matter field. The pattern of
soft supersymmetry breaking is also sensitive to this assignment, and would be
measurable at the LHC or in other collider experiments.

As we have mentioned, open issues in no-scale inflation include the mechanism for
stabilization of the various moduli, including the volume modulus that plays a prominent
r\^ole in building models. The inflationary observables are sensitive to the mechanism
of modulus stabilization, and therefore may be able to cast some light on this basic
issue in string phenomenology. More generally, for the foreseeable future measurements
of inflationary observables are likely to take us closer to the string scale than any other
experiments, and no-scale inflationary models may be the best platform for exploiting
this scientific opportunity.

\section*{Acknowledgements}

The work of J.E. was supported in part by the London Centre for Terauniverse Studies
(LCTS), using funding from the European Research Council via the Advanced Investigator
Grant 267352 and from the UK STFC via the research grant ST/L000326/1.
The work of D.V.N. was supported in part by the DOE grant DE-FG03-95-ER-40917 and in part by the Alexander~S.~Onassis Public Benefit Foundation.
The work of M.A.G.G. and
K.A.O. was supported in part by DOE grant DE-SC0011842  at the University of
Minnesota.

\end{document}